\documentclass{appolb}
\usepackage{graphicx}
\usepackage{amsmath}
\usepackage{amssymb}
\usepackage{cancel}
\usepackage{hyperref}
\usepackage{cite}
\usepackage{subfigure}

\def\msbar{\ensuremath{{\rm{\overline{MS}}}}} 
\newcommand{\der}{\ensuremath{{\operatorname{d}}}}
\newcommand{\GeV}{ {\rm GeV}}
\newcommand{\at}{\makeatletter @\makeatother}

\begin{document}
\title{Non-perturbative bottom PDF and its possible impact on new physics searches%
\thanks{Presented at the Cracow Epiphany Conference on the Physics in LHC Run2,
7-9 January 2016, Krak\'ow, Poland.}%
}
\author{Aleksander Kusina
\address{Laboratoire de Physique Subatomique et de Cosmologie\\ 
53 Rue des Martyrs Grenoble, France}
}
\maketitle
\begin{abstract}
Heavy quark parton distribution functions (PDFs) play an important role in several Standard Model
and New Physics processes. Most PDF analyses rely on the assumption that the charm and bottom PDFs
are generated perturbatively by gluon splitting and do not include any non-perturbative degrees
of freedom. However, a non-perturbative, intrinsic heavy quark PDFs have been predicted in the
literature. We demonstrate that to a very good approximation the scale-evolution of the intrinsic
heavy quark content of the nucleon is governed by non-singlet evolution equations, and use this
approximation to model the intrinsic bottom distribution and its impact on parton-parton
luminosities at the LHC.
\end{abstract}
%
  
\section{Introduction}
Heavy quark parton distribution functions (PDFs) play an important role
in several Standard Model (SM) and New Physics processes at the LHC. 
In particular, several key processes involve the bottom quark
PDF, e.g. $tW$, $tH^+$ production, associated $b$ plus $W/Z/H$ boson production
or $Hbb$ production~\cite{Maltoni:2012pa}.
In the standard approach employed by almost all global analyses of PDFs,
the heavy quark distributions  are generated {\em radiatively}, according
to DGLAP evolution equations 
starting with a perturbatively calculable boundary condition~\cite{Collins:1986mp,Buza:1996wv} 
at a scale of the order of the heavy quark mass.
In other words, there are no free fit parameters associated with the heavy
quark distribution and it is entirely related to the gluon PDF at the scale
of the boundary condition.
As a consequence, the uncertainties for the heavy quark and gluon
distributions are strongly correlated.

However, a purely perturbative, {\it extrinsic}, treatment where the heavy quarks are 
radiatively generated might {\em not} be adequate;
in particular for the charm quark with a mass $m_c \simeq 1.3$ GeV which is
not much bigger than typical hadronic scales but also for the bottom quark
with a mass $m_b \simeq 4.5$ GeV.
Indeed, there are a number of models that postulate a non-perturbative, {\it intrinsic},  heavy
quark component which is present even for scales $Q$ below the heavy quark mass $m$.
In particular, light-cone models predict a non-perturbative ('intrinsic') heavy quark
component in the proton wave-function~\cite{Brodsky:1980pb,Brodsky:1981se}
and similar expectations result from meson cloud
models~\cite{Navarra:1995rq,Melnitchouk:1997ig};
for a review of different models see e.g.~\cite{Brodsky:2015fna}.
Predictions of these models motivated people to investigate the possibility
of intrinsic charm (IC) in global PDF analyses~\cite{Pumplin:2007wg,Nadolsky:2008zw},
and gave the first estimate of how big the intrinsic charm could be.
Interestingly the two new global PDF analyses dedicated to IC from CTEQ~\cite{Dulat:2013hea}
and Jimenez-Delgado {\em et al.}~\cite{Jimenez-Delgado:2014zga} set significantly
different limits on the allowed IC contribution.%
    \footnote{This is partly because of the very different tolerance criteria
    which are used to define the range of acceptable fits.}

While there are at least a few global analyses 
which allow for an intrinsic charm component in the nucleon
\cite{Pumplin:2007wg,Nadolsky:2008zw,Dulat:2013hea,Jimenez-Delgado:2014zga},
studies of intrinsic bottom (IB) PDFs have not been performed at all.
In this contribution we summarize a technique, that we introduced in~\cite{Lyonnet:2015dca},
allowing to obtained IB (and IC) PDFs for any generic non-intrinsic PDF set.
Our approach exploits the fact that the intrinsic bottom PDF evolves (to an excellent precision)
according to a standalone non-singlet evolution equation and evolution of the other partons
is essentially not disturbed.

The rest of this paper is organized as follows.
In Sec.~\ref{sec:intrinsic}, we demonstrate that the scale-evolution
of the intrinsic PDF is governed by a non-singlet evolution equation. We then
propose suitable boundary conditions and perform
numerical tests of the quality of our approximations.
In Sec.~\ref{sec:numerics}, we use the IB and IC PDFs to obtain predictions
for parton--parton luminosities relevant at the LHC.
Finally, in Sec.~\ref{sec:conclusions}, we summarize our results and
present conclusions.
More details about this study can be found in~\cite{Lyonnet:2015dca}.

\section{Intrinsic heavy quark PDFs} 
\label{sec:intrinsic}

\subsection{Evolution equations}
\label{sec:evolution}
In the context of a global analysis of PDFs the different
parton flavors are specified via a boundary condition at
the input scale $\mu_0$ which is typically of the order
${\cal O}(1\ \GeV)$. Solving the DGLAP evolution equations
with these boundary conditions allows us to determine the PDFs
at higher scales $\mu > \mu_0$.
The boundary conditions for the up, down, strange quarks and
gluons are not perturbatively calculable and have to be determined
from experimental data. From this perspective, it is meaningless to
decompose the light quark and gluon PDFs
into distinct (extrinsic and intrinsic) components.
The situation is different for the heavy charm and bottom quarks 
where the boundary conditions have been calculated perturbatively.
A non-perturbative (intrinsic) heavy quark distribution $Q_1$ can then be
defined at the input scale $\mu_0$ as the difference of the full boundary condition for the
heavy quark PDF $Q$
and the perturbatively calculable (extrinsic) boundary condition $Q_0$:
\begin{equation}
Q_1(x,\mu_0):= Q(x,\mu_0) - Q_0(x,\mu_0)\, ,
\label{eq:Q1}
\end{equation}
where $Q=c$ or $Q=b$.
At NLO in the $\msbar$ scheme, the relation in Eq.~\eqref{eq:Q1} gets further simplified
if the input scale $\mu_0$ is identified with the heavy quark mass $m_Q$ because $Q_0(x,m_Q)=0$.
In this case, any non-zero boundary condition $Q(x,m_Q) \ne 0$ can be attributed to the intrinsic
heavy quark component.

Using the decomposition of Eq.~\eqref{eq:Q1} the DGLAP evolution equations
governing the scale dependence of PDFs can be written as%
    \footnote{Strictly speaking, the decomposition of $Q$ into $Q_0$ and $Q_1$ is 
    defined at the input scale where the calculable boundary condition for $Q_0$ is known. 
    Only due to the approximations in Eqs.\ \protect\eqref{eq:DGLAP2a} and \protect\eqref{eq:DGLAP2b}
    it is possible to entirely decouple $Q_0$ from $Q_1$ so that the decomposition becomes meaningful
    at any scale.} 
\begin{eqnarray}
\label{eq:DGLAP2a}
\dot g &=& P_{gg}\otimes g+P_{gq}\otimes q+P_{gQ}\otimes Q_0+ {\cancel{P_{gQ}\otimes Q_1}}\, ,
 \\
\label{eq:DGLAP2b}
\dot q &=& P_{qg}\otimes g+P_{qq}\otimes q+P_{qQ}\otimes Q_0+ {\cancel{P_{qQ}\otimes Q_1}}\, ,
 \\
\label{eq:DGLAP2c}
\dot Q_0 + \dot Q_1 &=& P_{Qg}\otimes g+P_{Qq}\otimes q+P_{QQ}\otimes Q_0+ P_{QQ}\otimes Q_1 \, .
\end{eqnarray}
Neglecting the crossed out terms which give a tiny contribution 
to the evolution of the gluon and light quark distributions the
system of evolution equations can be separated into two independent parts.
For the system of gluon, light quarks and extrinsic heavy
quark ($g,q,Q_0$) one recovers the standard evolution equation
without an intrinsic heavy quark component.
For the intrinsic heavy quark distribution, $Q_1$, one finds
a standalone non-singlet evolution equation
\begin{equation}
\dot Q_1 = P_{QQ}\otimes Q_1 \, . 
\label{eq:DGLAP4}
\end{equation}

To fully decouple the two evolution equations we need to allow for a violation
of the momentum sum rule. The violation is of the order of the momentum carried
by the intrinsic heavy quarks
\begin{equation}
\int_0^1\ \der x\  x\  \left(Q_1 + \bar{Q}_1\right)  
\end{equation}
which is known to be very small especially for the case of bottom quarks.

\subsection{Boundary condition}
\label{sec:bc}

The BHPS model~\cite{Brodsky:1980pb} predicts the following $x$-dependence
for the intrinsic charm (IC) parton distribution function:
\begin{equation}
c_1(x) = \bar c_1(x) \propto x^2 [6 x (1+x) \ln x + (1-x)(1+10 x+x^2)]\, .
\label{eq:bhps}
\end{equation}
However, the normalization and the precise energy scale of this distribution are
not specified.
In the CTEQ global analyses with intrinsic charm~\cite{Pumplin:2007wg,Nadolsky:2008zw}
this functional form has been
used as a boundary condition at the scale $Q = m_c$
and in this work we do the same.

In case of intrinsic bottom we expect that the $x$-shape of the
boundary condition will be very similar to that of intrinsic charm distribution.
However, the normalization of IB is expected to be parametrically 
suppressed by a factor $m_c^2/m_b^2 \simeq 0.1$.
Therefore, we propose the following boundary condition for the IB distribution
\begin{equation}
b_1(x,m_c) = \frac{m_c^2}{m_b^2} c_1(x,m_c)\, .
\label{eq:bc2}
\end{equation}

Let us also note that in our approach it would {\em not} be a problem to work with asymmetric
boundary conditions, $\bar c_1(x) \ne c_1(x)$ and $\bar b_1(x) \ne b_1(x)$,
as predicted for example by meson cloud models~\cite{Hobbs:2013bia}.

\subsection{Intrinsic heavy quark PDFs from non-singlet evolution}
\label{sec:PDFdef}

We have used approximation of Sec.~\ref{sec:evolution} and boundary conditions of
Eqs.~\eqref{eq:bhps} and~\eqref{eq:bc2} to produce a set of standalone IC and IB PDFs.
QCD parameters, such as the strong coupling or the quark masses were matched with
CTEQ6.6 fits~\cite{Nadolsky:2008zw}; normalization of IC PDF was fixed
to the one obtained in
CTEQ6.6c0 fit and IB normalization was respectively scaled.
Both PDFs were then evolved according to the non-singlet evolution
equation~\eqref{eq:DGLAP4} and corresponding grids were produced.

In order to test the ideas presented above
we use the CTEQ6.6c series of intrinsic charm fits
\cite{Nadolsky:2008zw}. 
The CTEQ6.6c series comprises 4 sets of PDFs including an intrinsic charm component.
Two of them, CTEQ6.6c0 and CTEQ6.6c1, employ the BHPS model with $1\%$ and $3.5\%$ IC probability, respectively.
This corresponds to the values of 0.01 and 0.035 of the first moment of the charm PDF,
$\int dx  \, c(x)$, calculated at the input scale $Q_0=m_c=1.3\ \GeV$.
In the rest of this contribution, we will follow the naming convention of the CTEQ6.6c fits
in which a given fit is characterized by the value in percentage of the first moment of
the charm distribution at
the input scale, e.g. 1\% for CTEQ6.6c0. 

In the following we compare our approximate IC PDFs supplemented with the central
CTEQ6.6 fit, which has a radiatively generated charm distribution, with the CTEQ6.6c0
and CTEQ6.6c1 sets where IC has been obtained from global analysis without the approximations
of Sec.~\ref{sec:evolution}.
\begin{figure}
\begin{center}
\includegraphics[angle=0,width=0.48\textwidth]{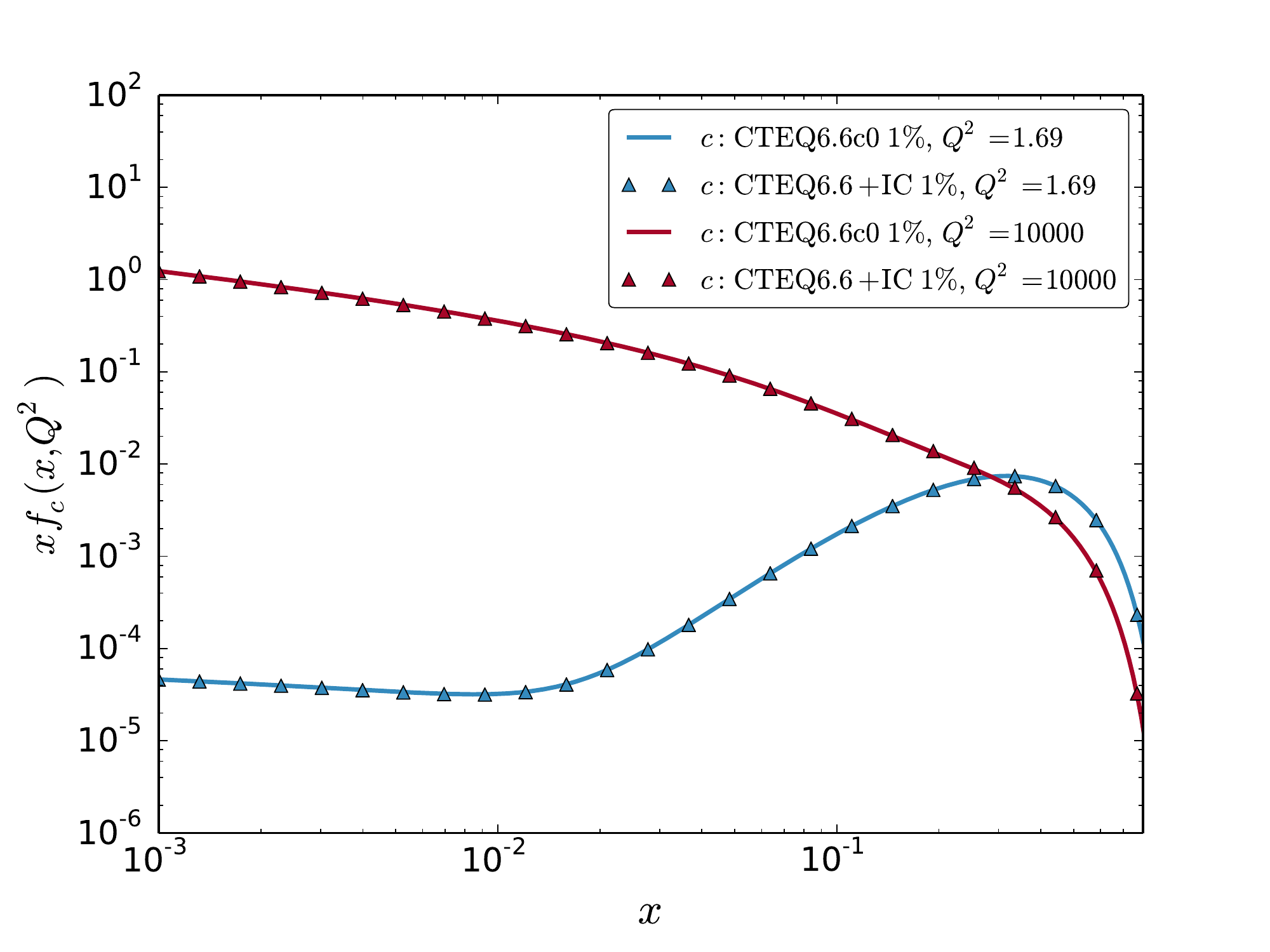}
\includegraphics[angle=0,width=0.48\textwidth]{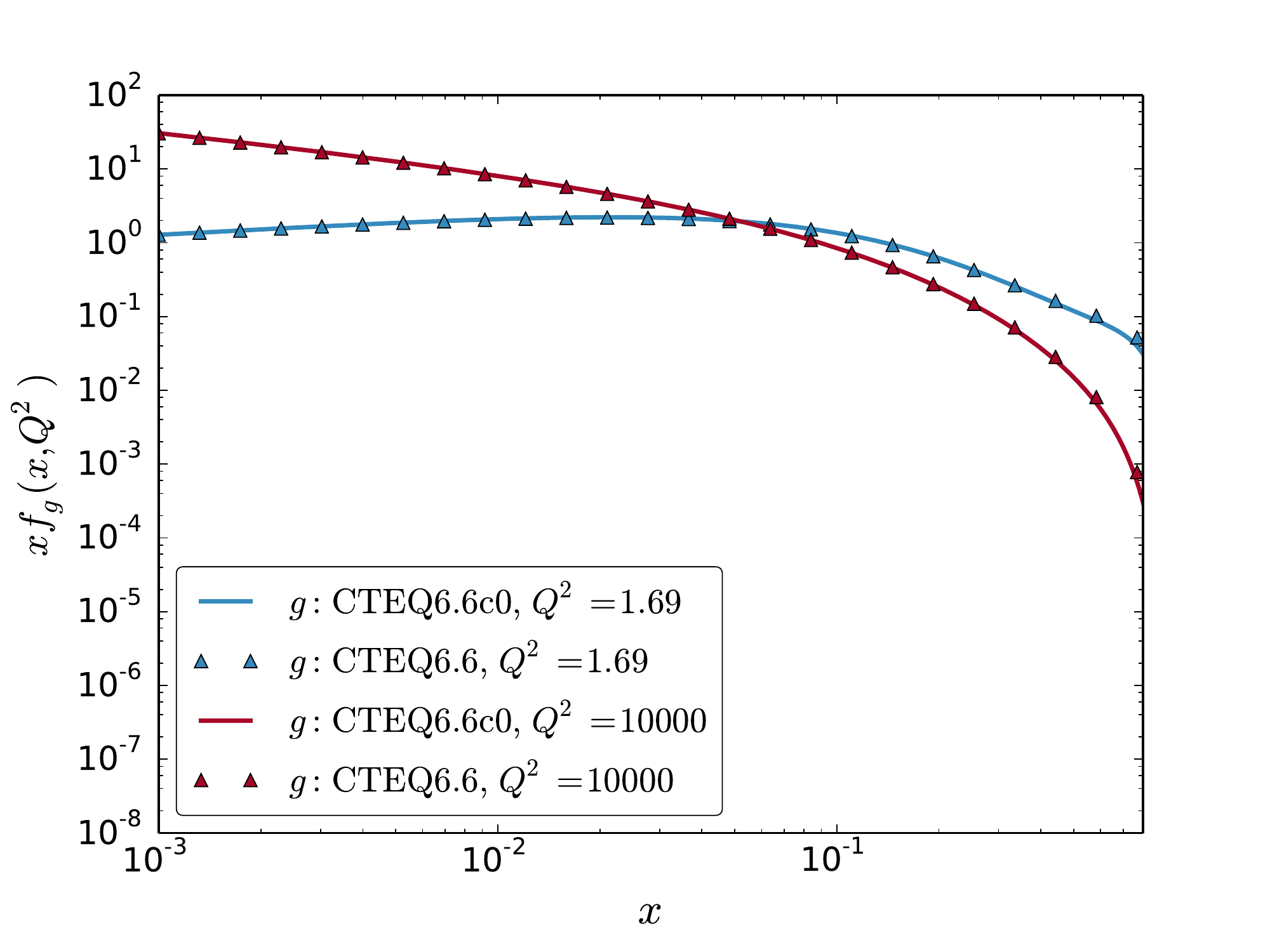}
\includegraphics[angle=0,width=0.48\textwidth]{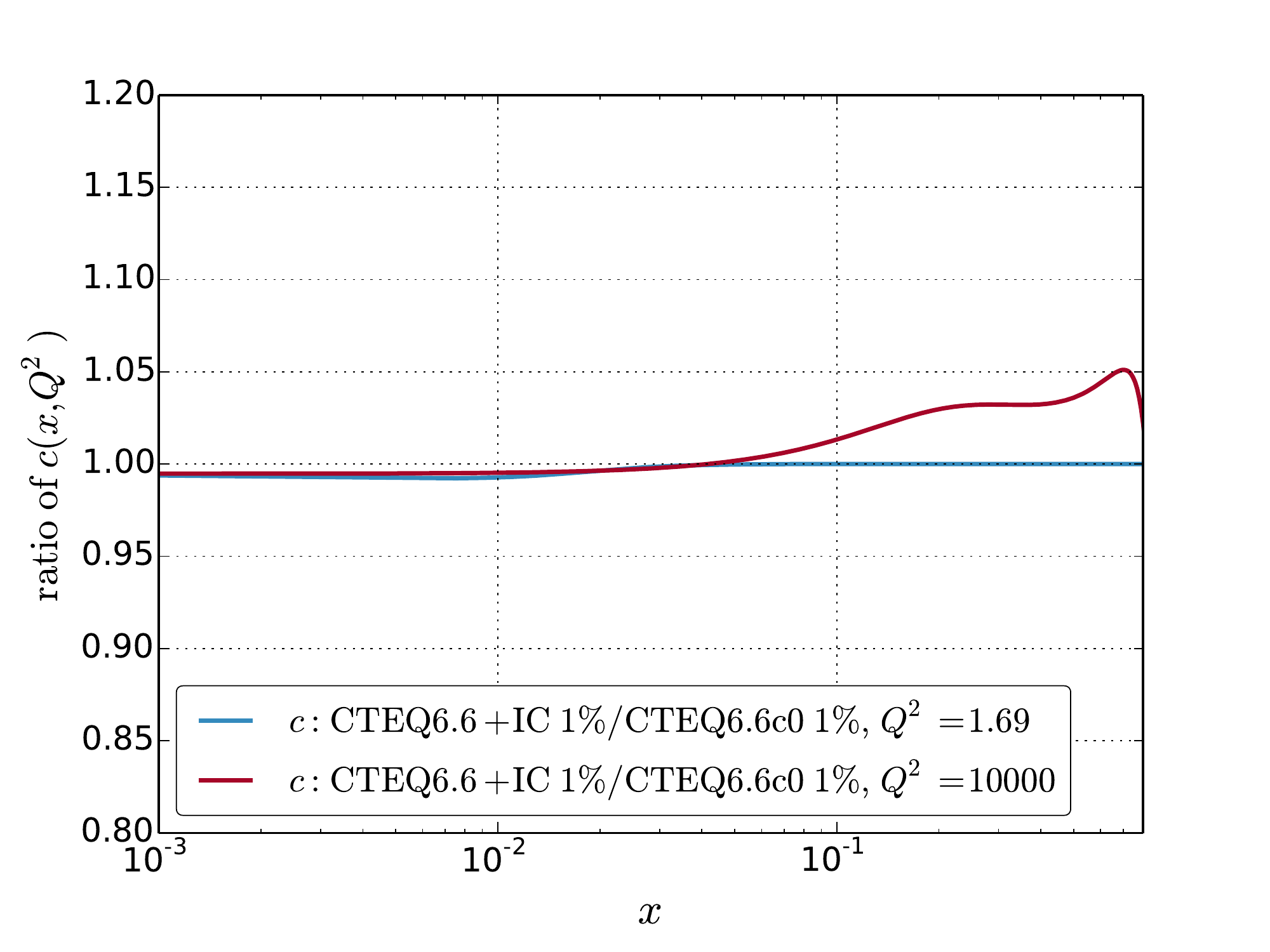}
\includegraphics[angle=0,width=0.48\textwidth]{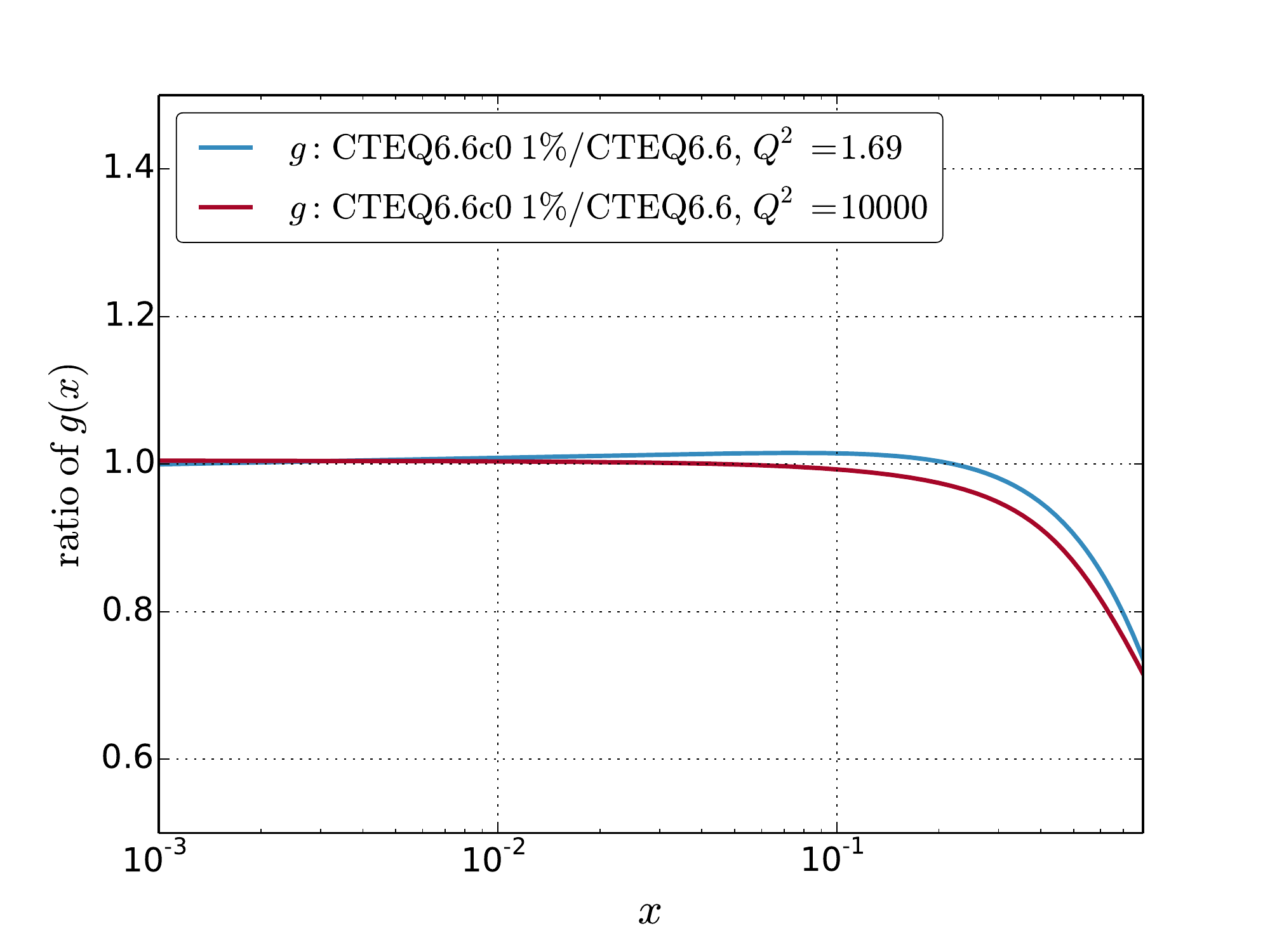}
\caption{
(upper left) CTEQ6.6c0 charm distribution function $c(x,Q^2)$ (solid lines)
and the sum $c_0(x,Q^2) + c_1(x,Q^2)$ (triangles) where $c_0$ is the radiatively
generated CTEQ6.6 charm distribution and $c_1$ is the non-singlet evolved IC.
(lower left) figure shows the ratio of the corresponding curves.
(upper right) Comparison of the CTEQ6.6c0 (solid line) and the CTEQ6.6 (triangles) gluon distributions.
(lower right) shows ratio of the  corresponding curves.
}
\label{fig:testa}
\end{center}
\end{figure}
In Fig.~\ref{fig:testa} (left panels) we present this comparison for 1\% IC normalization
for two scales $Q^2=1.69$ and $10000\ \GeV^2$.
As can be seen in the ratio plot the difference between the sum $c_0+c_1$ and the CTEQ6.6c0
charm distribution  is tiny at low $Q^2$, and smaller than $5\%$ at higher scales. 
In other words, the IC distribution $c_1$ evolved according to the decoupled non-singlet
evolution equation is in very good agreement with 
the difference $c-c_0$ representing the IC component in the full global analysis.

Of course the inclusion of the intrinsic charm distribution will 
alter the other parton distributions, most notably the gluon PDF.
In order to gauge this effect, in right panels of Fig.~\ref{fig:testa}, we compare gluon
distribution from the CTEQ6.6c0 analysis with the one from the standard CTEQ6.6 fit.  
For small $x$ ($x<0.1$) the gluon PDF is not affected by the presence of a BHPS-like intrinsic charm component
which is concentrated at large $x$.
At $x \simeq 0.7$, the CTEQ6.6c0 gluon is suppressed by about 20\% with respect to CTEQ6.6,
and this is relatively insensitive to the value of  $Q^2$. We note that at large-$x$, 
the gluon distribution is already quite small and the uncertainty of the gluon PDF is sizable
(of order of 40 -- 50\% for the CTEQ6.6 set).
The difference between the gluon distributions is {\em slightly} enhanced when evolving from the input scale $Q^2=1.69\ \GeV^2$
to the electroweak scale $Q^2=10000\ \GeV^2$, but it is still much smaller than the PDF uncertainty. 
We conclude that for most applications,  adding a standalone intrinsic charm distribution to an
existing standard  global analysis of PDFs is internally consistent and leads to only a small error.  
Moreover, for the case of intrinsic bottom which is additionally suppressed, 
the accuracy of the approximation will be even better. 

A more detailed numerical validation showing also effects on parton-parton luminosities has been
presented in ref.~\cite{Lyonnet:2015dca} where we have introduced this method.

\section{Possible effects of IC/IB on LHC observables}
\label{sec:numerics}

To provide a generic estimate of possible effects of IC and IB on LHC observables
we will investigate their impact on parton--parton luminosities at 14 TeV.
This allows us to assess the relevance
of a non-perturbative heavy quark component for the production of new heavy
particles coupling to the SM fermions.

Using the factorization theorem of QCD for hadronic cross sections, one can express
the inclusive cross section for the production of a heavy particle $H$ as follows: 
\begin{equation}
\sigma_{pp\rightarrow H+X} = \sum_{ij}  \int_\tau^1 \int_{\tau/x_1}^1 dx_1dx_2 f_i (x_1,\mu) f_j(x_2,\mu) \hat{\sigma}_{ij \rightarrow H}(\hat{s})\ ,
\label{eq:factorization}
\end{equation}
where $\tau=x_1 x_2 = m_H^2/S$, $S$ is the hadronic center of mass energy, and  $\hat{s}=x_1 x_2 S$ is its partonic
counterpart. $f_{i}(x,\mu)$ denotes the PDF of parton $i$ carrying momentum fraction
$x$ inside the proton. Finally, $\mu$ is the factorization scale which in the following is
identified with the partonic center of mass energy $\hat{s}=m_H^2$. Equation~\eqref{eq:factorization}
can be re-written in the form of a convolution of partonic cross-sections and parton--parton
luminosities,
\begin{equation}
\sigma_{pp\rightarrow H+X} = \sum_{ij} \int_\tau^1 d\tau \ \displaystyle \frac{\mathcal{L}_{ij}}{d\tau}\ \hat{\sigma}_{ij}(\hat{s}),
\label{eq:lumidef}
\end{equation}
where
\begin{equation}
	\frac{d\mathcal{L}_{ij}}{d\tau}(\tau,\mu) = \frac{1}{1+\delta_{ij}}\frac{1}{\sqrt{S}}\int_\tau^1 \frac{dx}{x}
                              \Big[ f_i(x,\mu)f_j(\tau/x,\mu) + f_j(x,\mu)f_i(\tau/x,\mu) \Big].
\label{eq:pplumi}
\end{equation}
All the results of this section have been obtained using the CTEQ6.6 PDF set~\cite{Nadolsky:2008zw}
supplemented with our approximate IC and IB PDFs constructed using the procedure presented
in Sec.~\ref{sec:intrinsic}.

\begin{figure}[h]
\begin{center}
\subfigure[]{
\label{fig:luminositiesa}
\includegraphics[width=0.48\textwidth]{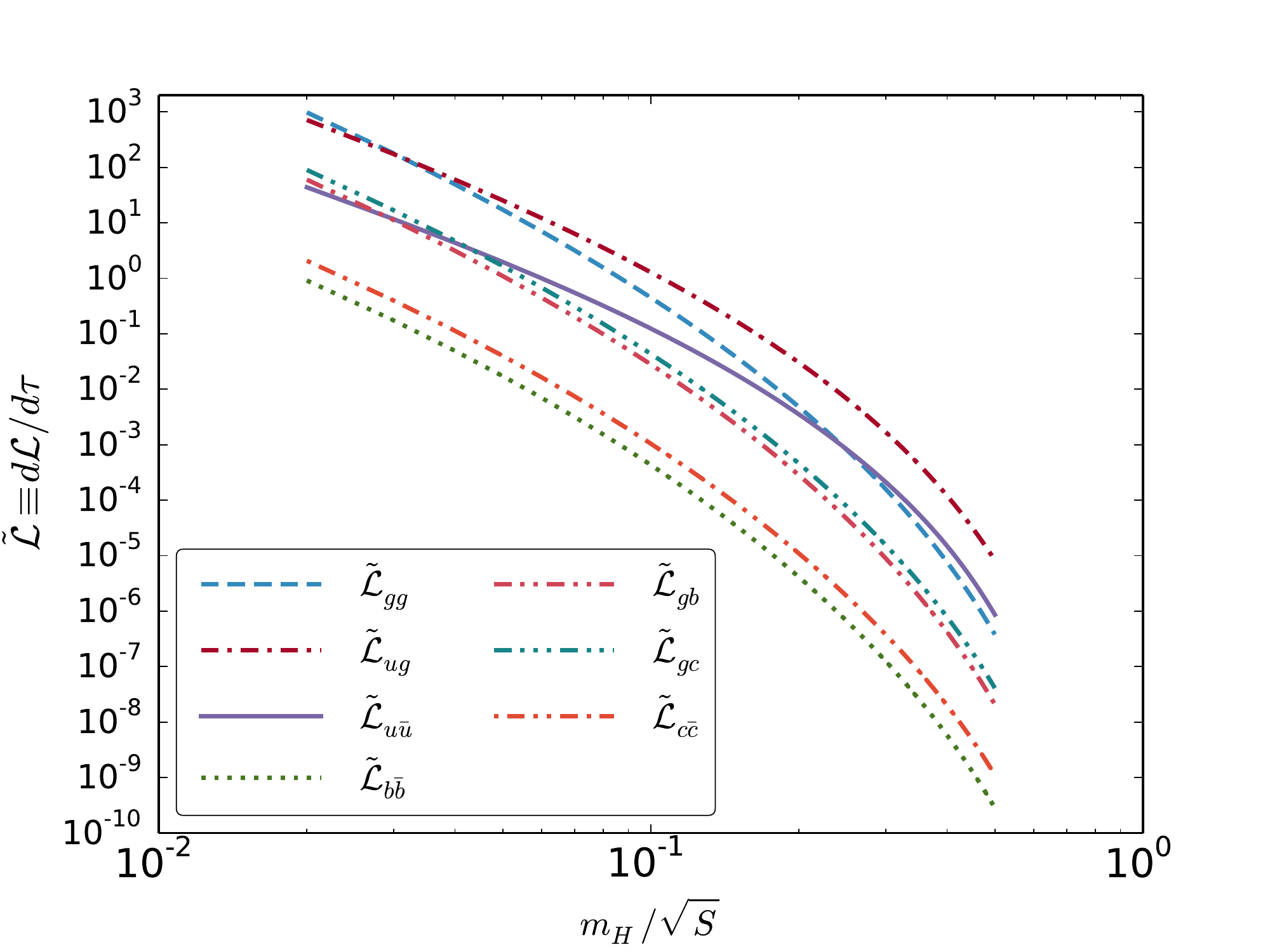}}
\subfigure[]{
\label{fig:luminositiesb}
\includegraphics[width=0.48\textwidth]{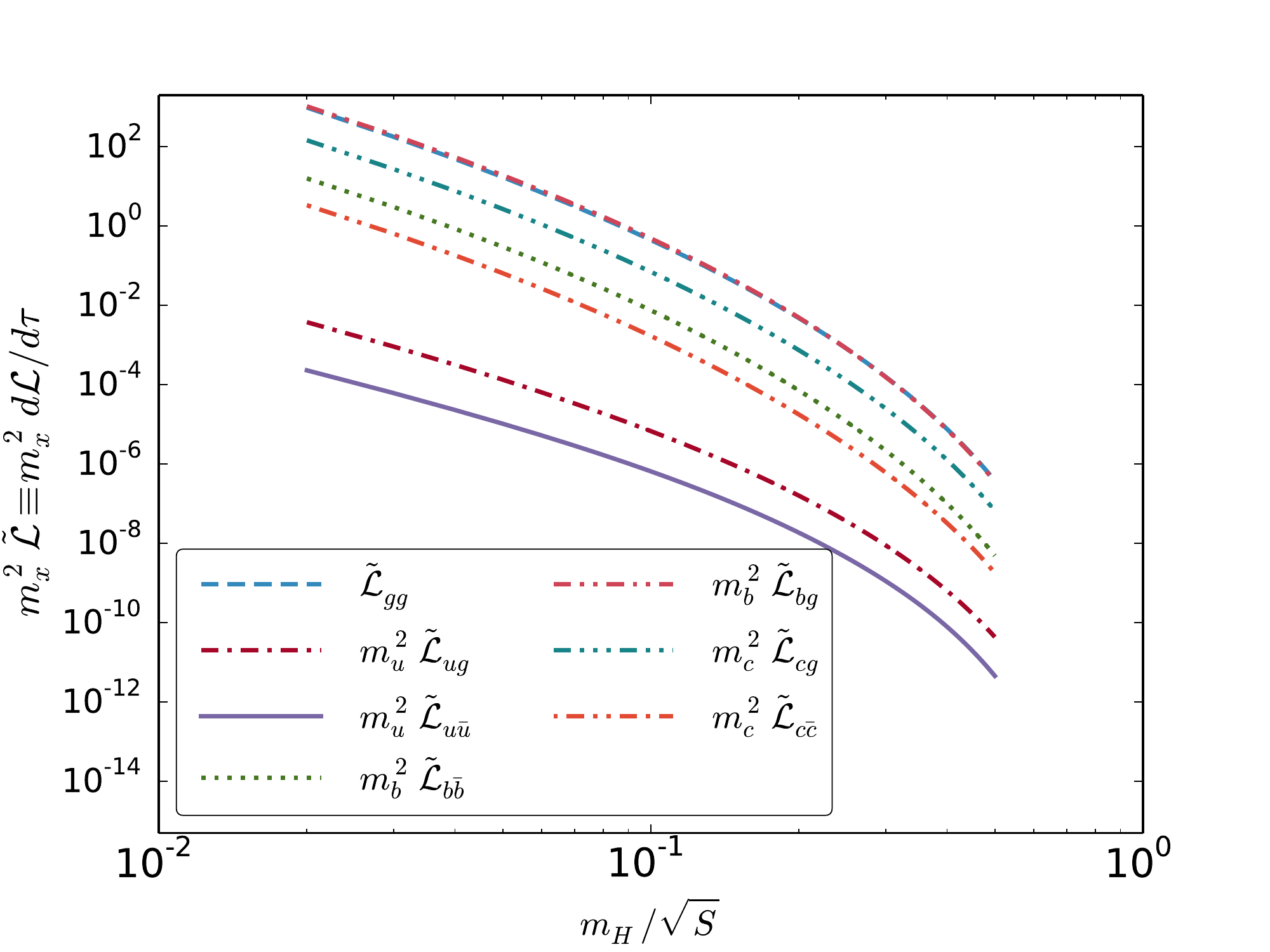}}
\caption{(a) Different parton--parton luminosities as a function of $\sqrt{\tau}=m_H/\sqrt{S}$
for the LHC14 calculated using CTEQ6.6 PDFs. 
For large $\tau$, the ordering of the curves is:
$ug \gg u \bar u > gg \gg gc > gb \gg c \bar c  > b \bar b$.
(b) Rescaled  
parton--parton luminosities ($m_i^2  d\mathcal{L}_{ij}/d\tau$) for the LHC14 calculated using CTEQ6.6 PDFs. 
For comparison, we also show the un-rescaled gluon--gluon luminosity.
For large $\tau$, the ordering of the curves is:
$gg \simeq gb > gc \gg b\bar b > c \bar c \gg u g \gg u \bar u$.
Note that by coincidence the gluon--gluon luminosity, $\mathcal{L}_{gg}$, agrees
at the 10\% level with the scaled $gb$ luminosity, $m_b^2 \mathcal{L}_{gb}$, so that the two curves
lie on top of each other in (b).
}
\end{center}
\end{figure}

In Fig.~\ref{fig:luminositiesa} we show different parton--parton luminosities, $d\mathcal{L}_{ij}/d\tau$,
for the LHC at 14 TeV (LHC14) as a function of $\sqrt{\tau}=m_H/\sqrt{S}$. We choose the range of $\sqrt{\tau}$
to be $[0.02,0.5]$ that corresponds to the production of a heavy particle of mass
$m_H\in[0.280,7]$ TeV which is roughly the range of values that will likely be probed at the LHC14.
As can be seen, at large $\sqrt{\tau}$, the parton--parton luminosities respect the following ordering:
$ug \gg u \bar u > gg \gg gc > gb \gg c \bar c  > b \bar b$.
Consequently, one can generally conclude that heavy quark initiated subprocesses play a minor role
in {\it most} processes where a heavy state is produced.

One exception would be SM extensions where the couplings to the first two generations
are suppressed or vanish so that the $gb$ or $b\bar b$ channels can dominate;
typically this is done in order to avoid experimental constraints from low energy
precision observables or flavor changing neutral currents. 
Of course, unless the couplings to the $gb$ or $b\bar b$ channels are enhanced, these scenarios
have tiny cross sections and will be difficult to measure at the LHC. 

However, if the couplings are enhanced by factors of the quark mass, 
the hierarchy of the contributions can change dramatically. 
This can happen when the heavy state has couplings to the
Standard Model particles proportional to their masses such as the SM Higgs or the Higgs particles in 2HDM models.
For example, in Fig.~\ref{fig:luminositiesa} we show the parton-parton luminosities 
with no enhancement factors; 
 in Fig.~\ref{fig:luminositiesb} we show the same but with additional factors proportional to the heavy quark mass;
the change is dramatic.
Taking the quark masses into account, the high $\tau$ region now exhibits the following hierarchy:
$gg \simeq gb > gc \gg b\bar b > c \bar c \gg u g \gg u \bar u$.
In this case the heavy quark initiated subprocesses could play the dominant role,
apart from the $gg$ initiated subprocesses which would contribute via an 
effective, model-dependent, heavy quark loop-induced $ggH$ coupling.

\begin{figure}[t]
\centering
\includegraphics[width=\textwidth]{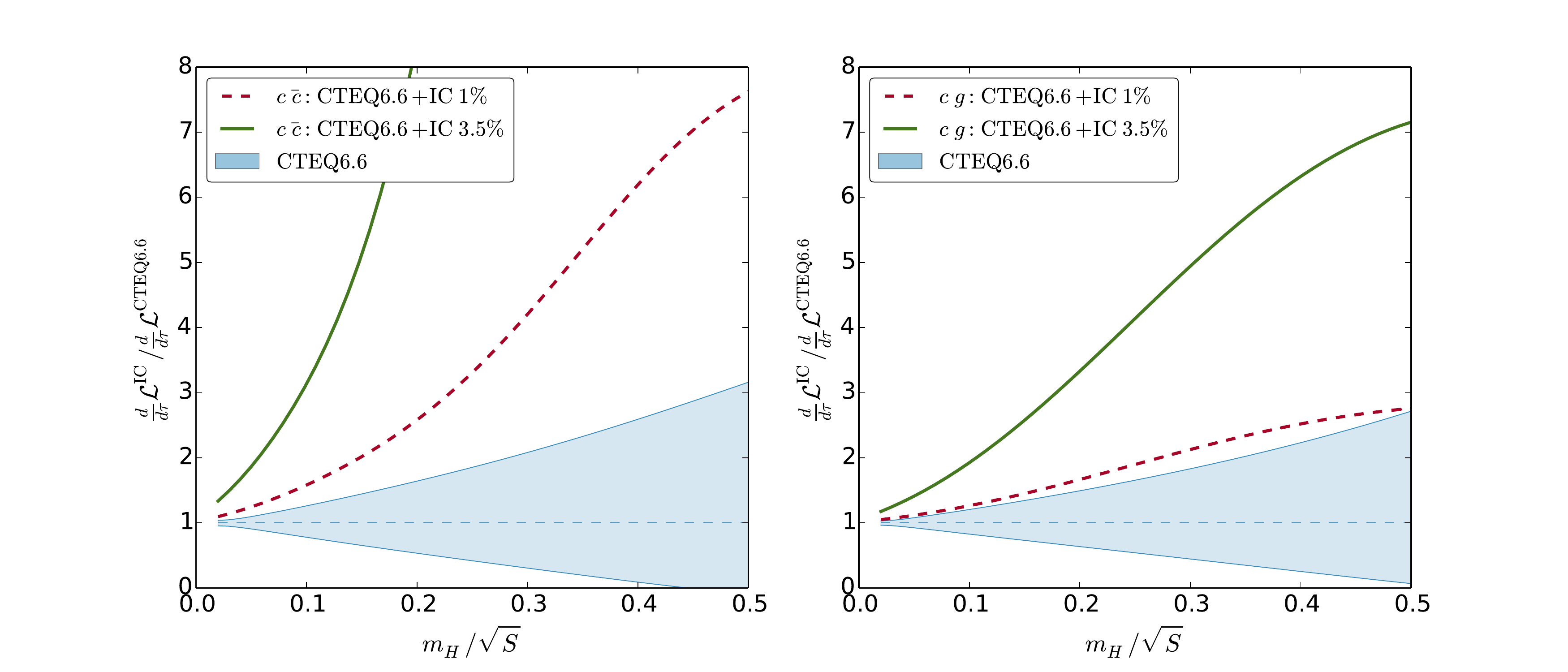}
\caption{Ratio of $c \bar c$ luminosities (left) and $c g$ luminosities (right) at the LHC14
for charm-quark PDF sets with and without an intrinsic component as a function of $\sqrt{\tau}=m_H/\sqrt{S}$. 
In addition to the curves with 1\% normalization (red, dashed lines) we include the results for
the 3.5\% normalization (green, solid lines).}
\label{fig:ratio_w_uncertainty_c}
\end{figure}

\begin{figure}[!h]
\centering
\includegraphics[width=\textwidth]{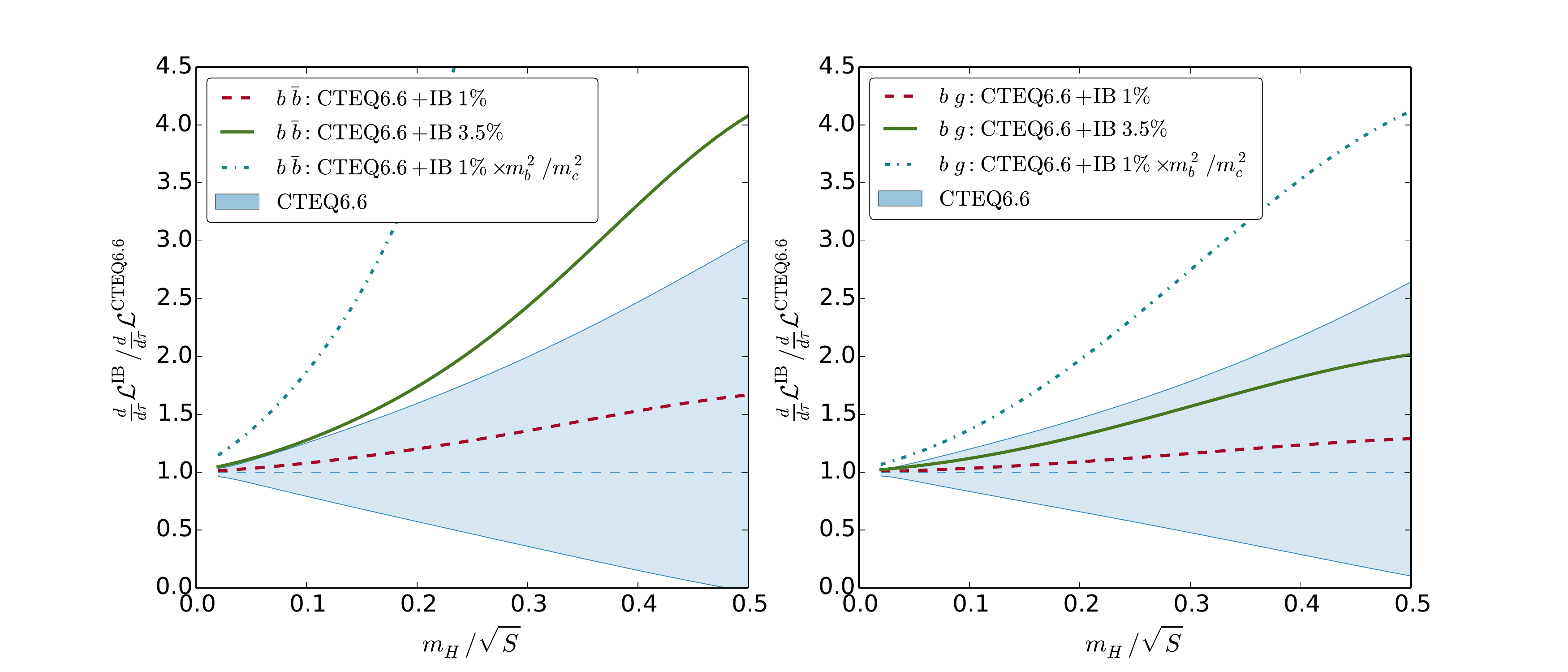}
\caption{Ratio of luminosities at the LHC14 for bottom-quark PDF sets with different
normalizations of the intrinsic bottom component. The plot has been truncated, and the $b\bar{b}$
luminosity in the extreme scenario reaches about 17 at $\sqrt{\tau}=0.5$.}
\label{fig:ratio_w_uncertainty_b}
\end{figure}

To explore how the presence of IC and IB would 
affect physical observables with a non-negligible heavy quark initiated subprocesses, 
in Figs.~\ref{fig:ratio_w_uncertainty_c} and~\ref{fig:ratio_w_uncertainty_b} we show
the ratios of luminosities for charm and bottom with and without an intrinsic contribution
for 1\% and 3.5\% normalizations.
Furthermore, since there are no experimental constraints on the IB normalization,
in Fig.~\ref{fig:ratio_w_uncertainty_b} we also include an extreme scenario where 
we remove the usual  $m_c^2/m_b^2$ factor; thus, the first moment of the IB is 1\% at the initial scale $m_c$.

For the 1\% normalization the  $c\bar c$ luminosity ratio 
grows as large as 7 or 8 for  $\sqrt{\tau}=0.5$, 
and for a 3.5\% normalization it
becomes extremely large and reaches values of up to 50.
From these figures we can clearly see that the effect of the 3.5\% IC is substantial
and can affect observables sensitive to $c\bar{c}$ and $cg$ channels. As expected, in the case of IB the effect is smaller but for the $b \bar{b}$ luminosity  the
IB with 3.5\% normalization leads to a curve which lies clearly above 
the error band of the purely perturbative result. 
In the extreme scenario (which is not likely but by no means excluded) the IB component has a big effect 
on both the $b\bar{b}$ and $bg$ channels.

%
\section{Conclusions}
\label{sec:conclusions}

We have presented a method to generate a matched IC/IB distributions
for any PDF set without the need for a complete global re-analysis. This allows one to easily carry
out a consistent analysis including intrinsic heavy quark effects. 
Because the evolution equation for the intrinsic heavy quarks decouples, 
we can freely adjust  the normalization of the IC/IB PDFs.

For the IB, our approximation holds to a very good precision.
For the IC, the error increases (because the IC increases), yet our method is still useful.
For an IC normalization of  1-2\%, the error is less than the  PDF uncertainties at the large-$x$ where the IC is relevant.
For a larger normalization, although the error may be the same order as the  PDF uncertainties, 
the IC effects also grow and can be separately distinguished from the case without IC. 
In any case, the IC/IB represents a non-perturbative systematic effect 
which should be taken into account.

The method presented here greatly simplifies our ability to search for, and place constraints upon, 
intrinsic charm and bottom components of the nucleon. This technique will facilitate
more precise predictions which may be  observed at future facilities such as an Electron Ion Collider (EIC), 
the Large Hadron-Electron collider (LHeC),  or AFTER\at LHC.

The PDF sets for intrinsic charm and intrinsic bottom discussed in this analysis
(1\% IC, 3.5\% IC, 1\% IB, 3.5\% IB) are available from the authors upon request.

\bibliographystyle{utphys}
\bibliography{refs}

\end{document}